\documentstyle[prd,eqsecnum,aps]{revtex}

\begin{document}

\draft

\title{
Radiation Damping in FRW Space-times with Dif\/ferent Topologies}  

\author{A. Bernui\cite{internet1},  G.I. Gomero\cite{internet2}, 
	M.J. Rebou\c{c}as\cite{internet3} and
		A.F.F. Teixeira\cite{internet4}  }

\address{Centro Brasileiro de Pesquisas F\'\i sicas \\
	 Departamento de Relatividade e Part\'\i culas \\
	 Rua Dr.\ Xavier Sigaud 150 \\
	 22290-180 Rio de Janeiro -- RJ, Brazil}

\date{\today}

\maketitle

\begin{abstract}
We study the role played by the compactness and the  
degree of connectedness in the time evolution of the energy of a 
radiating system in the Friedmann-Robertson-Walker (FRW) 
space-times  whose  $t=const\,$ spacelike sections 
are the Euclidean 3-manifold ${\cal R}^3$ and six topologically
non-equivalent
flat orientable compact multiply connected Riemannian 3-manifolds.
An exponential damping of the energy $E(t)$ is present in 
the ${\cal R}^3$ case, whereas for the six compact 
flat 3-spaces it is found basically the same pattern for the evolution 
of the energy, namely relative minima and maxima occurring
at dif\/ferent times (depending on the degree of connectedness)
followed by a growth of $E(t)$.
Likely reasons for this divergent behavior of $E(t)$  
in these compact flat 3-manifolds are discussed and
further developments are indicated.
A misinterpretation of Wolf's results
regarding one of the six orientable compact 
flat 3-manifolds is also indicated and rectified.
\end{abstract}
\vspace{5mm}

\section{Introduction}       \label{Intro}
\setcounter{equation}{0}

As general relativity is a purely metrical (local) theory it clearly
leaves unsettled the global structure (topology) of space-time. 
However, in cosmology perhaps the most important problems are related 
to the global structure of space-time, where the topological 
degrees of freedom ought to play an essential role.

Geometry constrains, but does not determine the topology of a
space-time. Consider, for example, the  Friedmann-Robertson-Walker
(FRW) space-times, whose line element can be given by 
\begin{equation}
ds^{2} =  dt^2 - A^2(t) \, \left[\, \frac{dr^2}{1 - \kappa\, r^2} +
r^2\,(\,d\theta^2\, + \, \sin^2 \theta \, d\varphi^{2}\,)\, \right]\,,
							\label{frwds2}
\end{equation} 
where $A(t)$ is the scale factor, $t$ is the cosmic time, and 
the constant spatial curvature  $\kappa = 0, \pm 1$ specifies the 
type of geometry (flat, elliptic or hyperbolic) of 
the $t=const$ spacelike section ${\cal M}_3$. Clearly FRW 
space-time manifolds ${\cal M}_4$ can be splitted into 
${\cal R} \times {\cal M}_3$.
The number of three-dimensional spacelike manifolds ${\cal M}_3$
which can be endowed with the three possible geometries of the $t=const\,$ 
3-spaces of FRW space-time is quite large~\cite{Wolf1967,Ellis1971}: 
for $\kappa = 0$ there are 18 topologically distinct 3-spaces, 
while for both $\kappa = \pm 1$ an infinite number of 3-spaces 
exist~\cite{Ellis1971,LachiezeReyLuminet1995}. 
Even if we restrict ourselves to orientable and compact manifolds
we still have an infinite number of 3-spaces ${\cal M}_3$ for the 
elliptic and hyperbolic cases, and six families of topologically 
dif\/ferent spacelike manifolds for the flat 
3-spaces~\cite{Wolf1967}~--~\cite{EllisSchreiber1986}.

Since physical laws are usually expressed in terms of local 
dif\/ferential equations, in order  to be confident about the 
physical results one derives it is often necessary to have some 
degree of control over the topological structure of the space-time 
manifold so as to include constraints imposed by the 
topology~\cite{OliveiraReboucasTeixeira1994}. 
One is then confronted with the question of what topologies are 
physically acceptable for a given space-time geometry. 
An approach to this problem is to study the possible 
observational (or physical) consequences of
adopting particular topologies for the 
space-time~\cite{Gott1980} -- \cite{Roukema1996}
(see also~\cite{LachiezeReyLuminet1995} and references therein).

In this work we study the role played by the compactness 
and the degree of connectedness in the 
time evolution of the energy of a radiating system 
in the flat FRW space-times whose spacelike 
$t=const\,$ sections are endowed with seven
different topologies, namely the simply connected 
3-space ${\cal R}^3$, and six multiply connected 
orientable compact 3-manifolds obtained by suitable identifications 
of opposite faces of cubes (four) and of hexagonal 
prisms (two) after suitable 
turns~\cite{Ellis1971,EllisSchreiber1986,Gomero1997}.

The radiating system we shall be concerned with is 
represented by a point-like harmonic oscillator (energy source)
coupled with a relativistic massless scalar field~%
\cite{SchwalbThirring1964}~--~\cite{Beig1978}.
Similar radiating systems have been used to study a wide 
class of radiation phenomena as, for example, 
gravitational waves~\cite{Unruh1983} or the radiated energy 
of oscillating electromagnetic dipoles~\cite{Jackson,Marion}.
Our radiating system was also used in
the study of dynamic and geometric constraints on the 
radiation in elliptic FRW expanding 
universes~\cite{Bernui1991}~--~\cite{Beig1978}.

In the next section the radiating system is described: the action 
integral is presented, its variation is performed and the 
corresponding evolution equations are obtained.
We also outline there a proposal for solving these equations.
In section~\ref{GreenFuns}  the Green functions 
of the wave  operator are derived 
for each specific space-time manifold with arbitrary
scale factor $A(t)$, and combined with the evolution
equations to give the corresponding radiation reaction equations.
A misinterpretation~\cite{Ellis1971}~--~\cite{EllisSchreiber1986}
of Wolf's \cite{Wolf1967} results (Theorem 3.5.5)
regarding one of the six orientable compact 
flat 3-manifolds is also indicated and rectified therein.
In section~\ref{NumAna}, the radiation reaction equations 
are numerically integrated. Graphs that show how 
the system energy varies with the time for each 
topologically dif\/ferent flat FRW manifold are 
presented for static and non-static cases.  
Our main conclusions are discussed in section~\ref{Conclusions}.
We show that when the $t=const$ sections are the simply connected
3-space ${\cal R}^3$ the radiation damping phenomenon is present,
whereas for all compact flat FRW space-time manifolds 
we have investigated, the energy $E(t)$ exhibits a few 
relative minima and maxima followed by a growth of the energy with 
the time.
Possible reasons for this divergent behavior of $E(t)$  
in these compact flat 3-manifolds are examined. 
We also discuss the role played by the compactness and by
the degree of connectedness in the energy patterns of our system 
in these space-time manifolds. 
Further developments are indicated therein.

\section{Physical System and Evolution Equations}
\label{SystEqs}
\setcounter{equation}{0}

Radiation waves produced by an oscillating energy source as,
e.g., an oscillating electromagnetic dipole, have been
studied by using a theoretical model represented by
a classical harmonic oscillator (energy source)
coupled with a relativistic massless scalar field
(scalar radiation waves propagating at speed of 
light)~\cite{SchwalbThirring1964}~--~\cite{Bernui1994}.
In our model the gravitational field is treated as external, 
but a suitable conformal coupling with the scalar field is 
considered~\cite{SonegoFaroni1993}.

The dynamics of our system can be described by an action integral, 
which contains a term for the scalar field $\phi(t,\vec{x})$, 
another associated to the oscillation amplitude $Q(t)$ of the
point-like harmonic oscillator, and a coupling term between the 
scalar field and the harmonic oscillator according to 
\begin{eqnarray} \label{action}
 S &=& \frac{1}{2} \int d^4x \, \sqrt{-g} \,\,[\, g^{\mu\nu}
  \partial_{\mu}\phi\,\partial_{\nu}\phi - \frac{1}{6} \hat{R}\,\phi^2 \,]
+ \frac{1}{2} \int dt \,[\,\dot{Q}^2 -\omega_{\varepsilon}^{2}\, Q^2\,]
\nonumber \\
{} &+& \lambda \int d^4x \, \sqrt{-g} \,\,\rho(t,\vec{x}) 
\,\,Q(t)\,\, \phi(t,\vec{x})\:, 
\end{eqnarray} 
where $t \in [t_0,\infty), \,\vec{x} \in {\cal M}_3$, $g_{\mu\nu}$ is 
the metric tensor on ${\cal M}_4$, $g \equiv \mbox{det}\,(g_{\mu\nu})\,$, 
$\hat{R}$ is the scalar curvature of ${\cal M}_4$, overdot means 
derivative with respect to $t$, and $\lambda$ is a coupling constant.
The function $\rho$ is the normalized density function, which
accounts for the coupling between the harmonic oscillator 
and the scalar field. Similarly to the coupling between
charges and electromagnetic fields in classical electrodynamics, 
we shall consider in this work a point-like coupling between
the harmonic oscillator and the scalar field, namely the one 
in which 
$\rho(t,\vec{x}) = \delta^{(3)}(\vec{x}) / \sqrt{-g(t,\vec{x})}$, 
where $\delta^{(3)}$ is the three-dimensional Dirac delta function.
This type of coupling requires a renormalization of the 
frequencies, and to this end we need  an $\varepsilon$-family of 
uncoupled frequencies $\omega_{\varepsilon}$ (see 
section~\ref{GreenFuns}). 
Here and in what follows units in which $c=1$ are used.

Varying the action~(\ref{action}) with respect to $\phi$ and $Q$ 
one obtains the coupled evolution equations of the system,
namely 
\begin{eqnarray}  \label{motion1}
\left[ \Box + \frac{1}{6} \hat{R}\, \right]\, \phi(t,\vec{x}) &=& 
\lambda \, \rho(t,\vec{x}) \, Q(t) \, ,
\\       \label{motion2}
\ddot{Q}(t) + \omega_{\varepsilon}^{2}(t) \, Q(t) &=& 
\lambda \int d^3x \,\sqrt{-g} \,\rho(t,\vec{x})\,\phi(t,\vec{x}),
\end{eqnarray}   
where $\Box \phi \equiv (\sqrt {-g})^{-1}\,\partial_\mu 
(\sqrt {-g} \:g^{\mu \nu} \partial_\nu \phi)$ 
is the  d'Alembertian operator, and $\Box + \frac{1}{6} \hat{R}\,$ 
is the wave operator defined on ${\cal M}_4$, 
hereafter simply called wave operator~%
\cite{SonegoFaroni1993,Friedlander1975}.

Before proceeding to the discussion of the Green functions 
for the wave operator in flat FRW space-time manifolds 
(for any smooth $A(t)$) we shall consider how one 
can obtain the radiation reaction equation from the above 
equations~(\ref{motion1}) and~(\ref{motion2}).
One first solves equation~(\ref{motion1}) as an initial 
value problem, by writing the solution in the form~%
\cite{MorseFeshbach1953} 
$\phi(t,\vec{x})=\phi_{I}(t,\vec{x})+\phi_{H}(t,\vec{x})$, 
where $\phi_{H}(t,\vec{x})$ satisfies the corresponding homogeneous 
equation, and where the solution of the inhomogeneous equation is 
given by
\begin{equation}   \label{inhoeq}
\phi_{I}(t,\vec{x}) = \int dt'\,d^3x'\, \sqrt{-g(t',\vec{x}')} \:\, 
{\cal G}(t,\vec{x};t',\vec{x}')\:\lambda\,\rho(t',\vec{x}')\,Q(t')\,,
\end{equation} 
with  $t' \in [t_0,t], \: \vec{x}' \in {\cal M}_3 $. In~(\ref{inhoeq})
${\cal G}(t,\vec{x};t',\vec{x}')$  is the retarded Green function, 
often referred to as fundamental solution of the 
wave operator~\cite{Friedlander1975}. 
Since the scalar field of our system may be thought of as 
the propagation medium for the radiating energy, we assume 
the initial condition
$(\phi(t_0,\vec{x}),\partial_{t}\phi(t_0,\vec{x})) = (0,0)$, which 
means that the scalar field carries no energy at $t = t_0$. This
condition implies that $\phi_{H}(t,\vec{x})=0$, and therefore 
$\phi(t,\vec{x})=\phi_{I}(t,\vec{x})$. 
Finally, the radiation reaction equation of the energy
source $Q(t)$ can be found by using the relation 
$\phi=\phi\,[Q]$ in equation~(\ref{motion2}). We shall 
return to this point later in the next section.

In this work, the Green function plays an essential role
in that it incorporates both the geometrical and 
topological features of the FRW $t=const\,$ 3-spaces. They 
will be obtained, in the next section, through the study of 
null geodesics of the space-times for 
each distinct 3-space (see table~\ref{FlatTops}).

\section{Green Functions for the Wave Operator} 
\label{GreenFuns}
\setcounter{equation}{0}

In this section we shall derive the Green functions of the 
wave operator for the flat FRW space-times whose spacelike
$t=const\,$ sections are the multiply connected compact
orientable 3-spaces, obtained by suitable identification of 
faces of a basic cell 
(see Refs.~\cite{Ellis1971,EllisSchreiber1986}) 
according to the table~\ref{FlatTops}, and the simply connected 
space ${\cal R}^3$.
Clearly the metric tensor we shall be concerned is 
given in cartesian coordinates 
by 
\begin{equation}
g_{\mu \nu} = \mbox{diag}\,(\,1,-A^2(t),-A^2(t),-A^2(t)\,)\,.
\end{equation} 

Regarding the manifold ${\cal T}_4$ one often encounters, in the 
literature on compact orientable flat 3-spaces, reference to a 
cube in which each pair of opposite faces is identified 
after half a turn%
~\cite{Ellis1971}~--~\cite{EllisSchreiber1986}.
However, if such a cube is endowed with the Euclidean geometry 
then the resulting three-space is not a manifold, but an
orbifold~\cite{Thurston1979,Thurston1982,Gomero1997}.
This 3-space fails to be a manifold along the edges of the 
cube.
Only when the cube is endowed with the elliptic geometry the 
resulting space is indeed a manifold, namely the well known real 
projective space ${\cal P}^3$. 

Recently, using theorem 3.5.5 by Wolf~\cite{Wolf1967}, 
it has been shown that a basic cell for the Euclidean manifold 
${\cal T}_4$ is the cube shown in figure 1~\cite{Gomero1997}.
It should be stressed that the basic cell for ${\cal T}_4$ shown 
in figure 1 is a standard basic cell; in general, however, the
basic cell for ${\cal T}_4$ needs not to be a cube.

%%%%
\begin{table}
\begin{center}
\begin{tabular}{||c |l |l||} \hline
{\bf Topology Type} & {\bf A Basic Cell} & {\bf Identifications of Faces} \\ 
\hline
\hline 
${\cal T}_1$ &  cube  &  3 pairs non rotated    \\ 
\hline
${\cal T}_2$ &  cube  &  2 pairs non rotated, 1 pair rotated $90^{\circ}$ \\ 
\hline
${\cal T}_3$ &  cube  &  2 pairs non rotated, 1 pair rotated $180^{\circ}$ \\ 
\hline
${\cal T}_4$ &  cube  &  1 pair rotated $180^{\circ}$, 2 pairs according to 
figure 1 \\ 
\hline
${\cal H}_1$ & hexagonal prism & top and bottom rotated $60^{\circ}$ \\ 
\hline
${\cal H}_2$ & hexagonal prism & top and bottom rotated $120^{\circ}$ \\
\hline
\end{tabular}
\end{center} 
\caption[]{\em The six compact orientable topologies for the 
flat 3-manifolds can be obtained by identifying opposite faces
of a basic cell as shown in this table.}
\label{FlatTops} 
\end{table}

\vspace{4mm} 
\noindent {\bf The cubes and $\bf {\cal R}^3$ cases} 
\vspace{2mm}

We shall discuss in this part the Green functions for the wave 
operator on 3-spaces ${\cal T}_i\,$ ($i = 1, \cdots ,4$) shown in 
table~\ref{FlatTops}. 
The manifold ${\cal T}_1$, referred to in the literature as 
the three-torus $T^3$~\cite{Weeks1985}, is a compact multiply 
connected Riemannian manifold, which can be obtained by 
identifying the opposite faces of a cube of side $a$. 

To build the Green function in ${\cal R} \times {\cal T}_{1}$
we first use the conformal properties of FRW space-times. 
Defining the conformal time 
$\tau \equiv f(t) \equiv \int dt'/ A(t')$ for $t' \in [t_0,t]$,  
we transform the problem of finding the Green function 
${\cal G}(t,\vec{x};t',\vec{x}')$ of the operator 
$(\Box + \frac{1}{6}\,\hat{R})$ 
into the problem of finding the Green function 
$G(\tau,\vec{x};\tau'=0,\vec{x}')$ of the operator 
$\Box \,$~\cite{AichelburgBeig1977,Friedlander1975}.
Moreover, a useful way of thinking about ${\cal T}_1$ in terms of 
the simply connected 3-manifold ${\cal R}^3$ is to imagine the cube 
repeated endlessly in a three-dimensional grid (basic cell and its 
images), where each repetition consists of the same physical region 
of space. 
The space ${\cal R}^3$ containing an infinite grid of copies of the 
basic cell is called the universal covering space of ${\cal T}_1$. 
In this way we now transform the problem of calculating the Green 
function of $\Box \,$ at $(\tau,\vec{x}) \in {\cal R} \times {\cal T}_{1}$ 
due to a point-like source located at $\vec{x}'=(0,0,0)$ and 
emitting at $\tau'=0$, to the equivalent problem of finding the field 
at $(\tau,\vec{x}) \in {\cal R} \times {\cal R}^{3}$ due to an infinite 
set of point-like sources, each one located at the center of a 
cube of the grid, and all of them irradiating simultaneously 
at $\tau'=0$. 

In this infinite grid picture the Green function of $\Box\,$ is 
calculated summing up the contributions of each one of the 
point-like sources. 
Since a scalar ray is a null geodesic, the distance travelled 
by each scalar wave front is a measure of the corresponding time 
travel, so the conformal time $\tau$ of arrival at $\vec{x}$ 
will depend on the position at which each source is located.
Null geodesics corresponding to retarded waves connecting 
$(\tau'=0,\vec{a}(\vec{n}))$ to the observation point $(\tau,\vec{x})$ 
are then such that $\tau - |\vec{x} - \vec{a}(\vec{n})| = 0$, so
the Green function for $\tau > 0$ is given by 
\begin{equation}  \label{greenT1}
G(\tau,\vec{x};\tau'=0,\vec{x}'=0)
= \frac{1}{4 \pi R(\tau)\, R(\tau'=0)} \; 
\sum_{n_{1},n_{2},n_{3}} \; 
\frac{\delta(\tau - |\vec{x} - \vec{a}(\vec{n})|)}
{|\vec{x} - \vec{a}(\vec{n})|} \; ,  
\end{equation} 
where $\vec{x},\, \vec{a}(\vec{n}) \in {\cal R}^3$,
$\vec{a}(\vec{n}) \equiv a \, (n_{1} \vec{e}_{1} 
+ n_{2} \vec{e}_{2} + n_{3} \vec{e}_{3})\,$. 
Here $\{ \vec{e}_{i} \}$ is the usual 
orthonormal basis vectors in ${\cal R}^3$ and 
$(n_{1},n_{2},n_{3})$ are integers, and 
$R(\tau) \equiv A(f^{-1}(\tau))$.

We shall consider now the radiation reaction equation
of our energy source in ${\cal T}_1$. Using the Green function
$G$ and (\ref{inhoeq}) one finds $\phi[Q]$, which can be used
in (\ref{motion2}) to furnish the radiation reaction equation
of our system.
Note that in these calculations the first term 
($n_{1} = n_{2} = n_{3} = 0$) 
of the Green function gives rise to the term
$ (\lambda^2/4\,\pi)\,\int d\chi\,\delta(\chi) / (A(t)\,\chi)$, 
that formally diverges. A suitable renormalization procedure is 
therefore necessary. This resembles the need for renormalization 
one finds when dealing with accelerated point charges in 
classical electrodynamics~\cite{Jackson}.
We learn from~\cite{Bernui1994} that a renormalization
procedure can always be made in this case. This amounts to saying
that for any $0 < \Omega^2 < \infty$, and any $\varepsilon > 0$, 
one can always define an $\varepsilon$-parameter family of 
frequencies $\omega_{\varepsilon}^{2}$ by
\begin{equation} \label{epsfam}
\omega_{\varepsilon}^2(t) \equiv \Omega^2 
+ \frac{2\,\Gamma}{A(t)} \int d\chi 
\frac{\rho_{\varepsilon}(\chi)}{\chi}\,,
\end{equation} 
such that $\rho_{\varepsilon}(\chi)$ is a $\delta$-sequence that 
in the limit $\varepsilon \rightarrow 0$ converges to the Dirac 
delta $\delta(\chi)$. 

Using now (\ref{epsfam}), in which we have 
made $2\,\Gamma \equiv \lambda^2/(4\pi)\,$, one finds that, 
for arbitrary initial data $(Q(t_0),\dot{Q}(t_0))$,
the renormalized radiation reaction equation of the source 
can be written in the  form
\begin{equation}  \label{T1radeq}
\ddot{Q}(t) + 2\, \Gamma\, \dot{Q}(t) + \Omega^2 Q(t) =  
\frac{2\, \Gamma}{A(t)} \, 
 \sum_{n_{1},n_{2},n_{3}} \frac{1}{a(\vec{n})}
 \Theta(t-t_n) \, Q(f^{-1}(f(t) - a(\vec{n})\,)) \,,
\end{equation} 
for any $t \in [t_{0},\infty) \,$, and where 
$a(\vec{n}) \equiv |\vec{a}(\vec{n})| 
= a \, [ {n_{1}}^2 + {n_{2}}^2 + {n_{3}}^2 ]^{1/2}$ 
for all $(n_{1},n_{2},n_{3})$ integers not all zero,
$t_n \equiv f^{-1}(a(\vec{n}))\,$ and the step distribution 
$\Theta(s) = 0$ for any $s \leq 0$, 
$\Theta(s) = 1$ for all $s > 0$.
Note that the radiation reaction equation 
(\ref{T1radeq}) contains an infinite, but  countable, 
number of retarded terms. Note also that, for a given pair
of initial data ($Q(t_0), \dot{Q}(t_0)$\,), the continuity 
of $Q$ and $\dot{Q}$ for all $t$ is suf\/ficient 
to ensure that (\ref{T1radeq}) can be integrated to give 
a unique solution. 
 
The next 3-manifolds we shall consider are ${\cal T}_2$, ${\cal T}_3$ 
and ${\cal T}_4$ (see table~\ref{FlatTops}). The 3-space ${\cal T}_2$ 
is a compact multiply connected flat three-dimensional Riemannian 
manifold, obtained by identifying opposites faces of a cube, with a 
pair being identified after a rotation of $90^\circ$  of
one face relative to the opposite face. 
The 3-manifold ${\cal T}_3$ is obtained by 
identification of opposites faces of a cubic cell, 
but now a pair of faces are identified 
after a turn of $180^\circ$ (see table~\ref{FlatTops}).
Finally, the 3-manifold ${\cal T}_4$ is obtained by 
pairwise identifying the faces of the cube according to figure 1.
For these 3-spaces the Green functions for the wave operator 
can again be built by using the infinite grid of cubes picture. 
It should be noticed, however, that the grids 
of cubes corresponding to $\;{\cal T}_1$, $\,{\cal T}_2$, 
$\,{\cal T}_3\,$ and $\,{\cal T}_4\,$ are dif\/ferent
\footnote{This is so because each 3-manifold is obtained by 
forming the quotient 
${\cal M}_3 = {\cal R}^3 / \Gamma_i$ ($i=1, \cdots ,4$),
where for each case $\Gamma_i$ is a dif\/ferent discrete 
group of isometries of the covering space ${\cal R}^3$ 
acting properly discontinuously,
without fixed points~\cite{Ellis1971}.}.
Nevertheless, as the center of a basic cell in each case 
represents the same infinite set of points (images of the 
point-like source at ($0,0,0$) ) of the covering space 
${\cal R}^3$, the Green functions for the wave operator in 
these 3-spaces are equal to the one we have calculated for 
${\cal T}_1$, i.e., are given by (\ref{greenT1}).
Thus, for arbitrary initial data $(Q(t_0),\dot{Q}(t_0))$
the radiation reaction equation in  ${\cal T}_2$, ${\cal T}_3$ 
and ${\cal T}_4$ reduces to (\ref{T1radeq}) obtained for 
${\cal T}_1$.

A word of clarification is in order here: the above Green 
functions for the wave operator for the 3-spaces ${\cal T}_i$ 
($i=1,\cdots,4$) coincide only because in each case the point-like 
source is located at the center of the basic cell $\vec{x}' = 0$.
This is so because although all these 3-spaces are locally 
homogeneous, 
the 3-spaces ${\cal T}_2$, ${\cal T}_3$  and ${\cal T}_4$ 
are not globally homogeneous, thus the corresponding Green functions 
of $\Box \,$ for $\vec{x}'\not=0$ are more involved and do not 
depend simply on the relative position between the source
and the observer.

Regarding the simply connected flat 3-dimensional Riemannian 
manifold ${\cal R}^3\,$ ($t=const\,$ section of the Minkowski 
space-time), the Green function for the wave operator $\Box$
is equal to the first term ($n_{1}=n_{2}=n_{3}=0$) of $G$ 
appearing in (\ref{greenT1}). 
Thus the radiation reaction equation in this manifold reduces 
to (\ref{T1radeq}) with right hand side equal to zero, for
all $t \in [t_{0}, \infty)$. 

\vspace{4mm} 
\noindent {\bf The hexagonal prism cases} 
\vspace{2mm}

We shall discuss now the Green functions for the wave operator
on the flat hexagonal prism manifolds ${\cal H}_1$ and 
${\cal H}_2$ described in table~\ref{FlatTops}. 

The 3-space ${\cal H}_1$ is a compact multiply connected flat 
three-dimensional Riemannian manifold, which can be obtained 
by identifying the top and bottom faces of a regular hexagonal 
prism after a rotation of $60^{\circ}$, while the lateral faces 
are pairwise identified in the usual manner.
The 3-manifold ${\cal H}_2$ can be similarly defined, but
now the identification of the top and bottom faces 
(regular hexagons) is made after a turn of $120^{\circ}$.
In both cases we denote by $a$ the shortest distance between two 
opposite sides of the regular hexagon and by $h$ the height of 
the hexagonal prism cell. 

For a point-like source located at the center of the basic cell, 
the Green functions for the wave operator in the 3-manifolds 
${\cal H}_1$ and ${\cal H}_2$ are obviously the same. 
Moreover they turn out to be similar to the Green 
function~(\ref{greenT1}). But now instead of $\vec{a}(\vec{n})$ 
one has $\vec{b}(\vec{n}) \equiv \, 
  a\, n_{1} \frac{\sqrt{3}}{2} \vec{e}_{1} 
+ a\, ( \frac{1}{2} n_{1} + n_{2} ) \vec{e}_{2} 
+ h\, n_{3} \vec{e}_{3}\,$, where  $(n_1,n_2,n_3)$ are integers, 
to locate the centers of the hexagonal prism cells in which 
${\cal R}^3$ has been tessellated.
Note that the number of images at a given distance from the
point-like source clearly depends upon the ratio $h/a$. 
The radiation reaction equation in the present cases 
is similar to~(\ref{T1radeq}) but with the obvious change 
$a(\vec{n}) \rightarrow b(\vec{n}) \equiv 
|\vec{b}(\vec{n})| = 
[ a^2 ( {n_{1}}^2 + n_1 n_2 + {n_{2}}^2 ) + h^2 {n_{3}}^2 ]^{1/2}$ 
and again the integers $(n_1,n_2,n_3)$ are not all zero. 

It should be emphasized that as the 3-manifolds 
${\cal H}_1$ and ${\cal H}_2$ are locally but not globally
homogeneous, the Green function for the wave operator in
these manifolds is again given by~(\ref{greenT1}), with $\vec{b}(\vec{n})$ 
instead of $\vec{a}(\vec{n})$, only because the point-like source is
at the center of the basic cell $\vec{x}'=0$.

To close this section we emphasize that the Green functions, 
obtained for flat FRW space-time manifolds 
${\cal M}_4 = {\cal R} \times {\cal M}_3$, contain the 
topological constraints of ${\cal M}_3$, information that makes 
possible to find out the exact radiation reaction equation in each 
case.

\section{Numerical Analysis}
\label{NumAna}
\setcounter{equation}{0}

In this section we shall discuss the time behavior of the energy 
of the harmonic oscillator 
$E(t) = \frac{1}{2}\,[\,\dot{Q}^{2}(t) + \Omega^2\, Q^{2}(t)]$, 
where the function $Q(t)$ is the solution of the radiation 
reaction equation corresponding to each flat manifold we have 
discussed in the previous section. 

Without loss of generality, in the integration of the 
radiation reaction equations and in the 
plotting of the energy function we have taken specific
values for the constants. We have also assumed the continuity of
$Q$ and $\dot{Q}$, and chosen suitable
values for the initial data $Q(t_{0})$ and $\dot{Q}(t_{0})$.
For a neat comparison between the simply and multiply connected
cases we have chosen  $\Gamma = 1$, $\Omega^2 = 30 $, the length 
$a=1$, and for the heights of the hexagonal prism and three-torus 
we have taken $h =0.4$ and $1$.
As a matter of fact, our three-torus ${\cal T}_1$ was obtained 
from a parallelepiped with edges $a,\, a,\, h$, so 
$a(\vec{n}) = 
[ a^2 ( {n_{1}}^2 + {n_{2}}^2 ) + h^2 {n_{3}}^2 ]^{1/2}$ 
was used in equation (\ref{T1radeq}). 
Further, we have also chosen as initial data $t_0=0$, 
$(Q(0),\dot{Q}(0)) = (\sqrt{2} / \Omega, 0)$, 
which means that the initial energy of the source has been 
normalized, i.e., $E(0)=1$.

Figures 2, 3 and 4 correspond to the static case 
$A(t)=const\, \equiv A_0$, which we have normalized to 1 
($A_0=1$), while dynamic situations ($\dot{A}(t) \neq 0$) are 
considered in figure 5. 

Taking into account the above choices of values and
using the computer algebra systems 
{\em Mathematica}~\cite{Wolfram1991} and {\em Maple}~\cite{Maple} 
the numerical integrations of the radiation reaction equations 
as well as the corresponding graphs for the energy function
were obtained (see figures 2 to 5). 

Figure 2 shows the behavior of the energy with
the time for the flat FRW space-times, in which the $t=const\,$
sections are any of the orientable compact 
multiply connected 3-manifolds 
${\cal T}_1$, ${\cal T}_2$ and ${\cal T}_3$ and ${\cal T}_4$ 
for dif\/ferent ratios 
$h/a$. Note that $h$ is the height of the parallelepiped 
of square basis of side $a$. 
The curves exhibit basically the same pattern, namely relative
minima and maxima followed by a predominant growth of the energy 
with the time ($E(t) \rightarrow \infty $ when $t \rightarrow \infty$). 
These extrema are related to the contribution of the discrete 
retarded terms (the right hand side of the radiation reaction 
equation) demanded by the compactness and the corresponding 
connectedness of these manifolds. 
The fact that the relative extrema occur at dif\/ferent 
times and are of dif\/ferent amplitudes (intensities) for distinct
tessellations (dif\/ferent ratios $h/a$) of the covering manifold, 
basically reveals the dif\/ferences in their degree of connectedness 
(returning rays take dif\/ferent times to return to 
the origin). This growth of the energy is discussed in 
the next section. 

Regarding the behavior of the energy function for the cases 
in which the $t=const$ sections are ${\cal H}_1$ and 
${\cal H}_2$, shown in figure 3, we again note that due to the 
compactness and connectedness we have relative minima and maxima
with distinct intensities, which take place at dif\/ferent 
instants for distinct ratios $h/a$.
The curves again display an eventual growth of the energy with 
the time for these manifolds.

Figure 4 compares the variation of the energy $E$ with the 
time for the cases where the $t=const\,$ section 
${\cal M}_3$ are ${\cal R}^3$, ${\cal T}_1$ (with $a=h=1$) 
and ${\cal H}_1$ (also with $a=h=1$). 
This figure shows for the Minkowski space-time, as expected, 
an exponential decay of the energy with the time, whereas for the 
manifolds ${\cal T}_1$ and ${\cal H}_1$ it shows basically the 
same pattern, i.e. relative minima and maxima occurring at dif\/ferent 
times, depending on the degree of connectedness, followed
by a growth of the energy with the time.

Although the net role played by the degree of
connectedness as well as compactness can be singled out in the 
static cases $A(t)= const\, \equiv A_0$, for the sake of 
completeness we have examined three instances where dynamic 
expansion takes place. Figure 5 corresponds to the plot of the 
energy function for the ${\cal T}_1$ manifold (with $a=h=1$) in 
the dynamic expanding cases:
(i) linear expansion $A(t) = t + 0.7$, 
(ii) square root expansion $ A(t) = \sqrt{t}$, 
(iii) inflationary expansion $A(t) = e^{t/\sqrt{2}}$, 
that is, an expansion with future event horizon.

Although for the case (iii) one clearly has radiation damping,
we have not been able to find out so far a closed formal 
proof of the asymptotical behavior for the other two cases.
We emphasize, nevertheless, that the net role played by the 
connectedness and compactness can be better singled out 
in the static cases, where the dynamical degrees of freedom
are frozen. The study of that role for the static cases 
is in fact the major aim of the present work.

\section{Conclusions and Final Remarks}
\label{Conclusions}
\setcounter{equation}{0}

In this work we have studied the role played by the topological
compactness and connectedness in the time evolution of the energy
of an harmonic oscillator in flat FRW space-time manifolds, whose 
$t=const\,$ sections are 
(i) the orientable simply connected non-compact 3-space ${\cal R}^3$ 
and (ii) six possible flat orientable multiply connected compact 
3-manifolds given in table \ref{FlatTops}.

For the ${\cal R}^3$ case we found that the energy 
function $E(t)$ exhibits an exponential 
decay with the time --- the radiation damping 
($E(t) \rightarrow 0$ when $t \rightarrow \infty$) takes place, 
as one could have  expected in agreement with~%
\cite{SchwalbThirring1964}~--~\cite{HoenselaersSchmidty}.

For the manifolds ${\cal T}_1$, ${\cal T}_2$, ${\cal T}_3$ and
${\cal T}_4$ as well as for the manifolds ${\cal H}_1$ and 
${\cal H}_2$ the behavior of the energy with the time exhibits 
basically the same pattern:
relative minima and maxima occur at dif\/ferent times
for distinct ratios $h/a$ (distinct tessellations) depending on
the degree of connectedness of each 3-manifold, and are 
followed by a growth of $E(t)$.

This asymptotical divergent behavior of $E(t)$ for these
compact manifolds contrasts with the radiation damping of the energy 
we have found for ${\cal R}^3$. There is a quite simple
heuristic argument which suports our numerical results, though.
If one ad hoc assumes an exponential asymptotical behavior
for $Q$, i.e. $Q(t) = \gamma\,\, \mbox{exp}\,(\beta \, t)$ with
$\beta$ and $\gamma$ real constants, then for the static cases ($A(t)=1$),
and for each of the above compact manifolds, in the limit
$t \rightarrow \infty$ equation~(\ref{T1radeq}) reduces to
\begin{equation}  \label{Einfty}
 \frac{\beta^2 + 2\, \Gamma\, \beta + \Omega^2}{2\,\Gamma} =
\sum_{m=1}^\infty \; \frac{c_m}{a_m}\,\, \exp (-\beta\,a_m) \; ,  
\end{equation} 
where $c_m$ is the number of images of the point-like source at
a distance $a_m$ in the infinite grid picture. To attain our
goal, we will show that~(\ref{Einfty}) has only 
one real solution for $\beta$, which is positive. 
Indeed, let $f(\beta)$ be the right hand
side of~(\ref{Einfty}), which is a positive monotone decreasing
function of $\beta$, and such that
\begin{equation}
\lim_{\beta \to  0} \: f(\beta) = \infty  \qquad \quad
\mbox{and} \qquad \quad
\lim_{\beta \to  \infty } \: f(\beta) = 0 \,. 
\end{equation}
Thus $f(\beta)$ lies entirely in the first quadrant of the plane
and crosses it from the top-left to the bottom-right.
Now, since $\Gamma > 0$ then for a given pair ($\Gamma, \Omega$)
the left hand side of (\ref{Einfty}) is a parabola curved upwards
with vertex at $\beta = - \Gamma$. Therefore, it always intersects
the curve for $f(\beta)$ in just one point, which is in the first 
quadrant. 
In other words, there is only one real $\beta$ solution to 
equation~(\ref{Einfty}), which is positive.

The unexpected (unphysical ?) growth of the energy with the time 
for the above compact flat 3-manifolds cases illustrates that non-trivial 
topologies can induce rather important dynamic changes in the
behavior of a physical system. This type of sensitivity has
been refered to as {\em topological fragility\/} and can occur 
without violation of any local physical law%
~\cite{ReboucasTavakolTeixeira1996}. 
A rigorous {\em non-numerical\/} analysis of the 
reasons for this surprising divergent behavior of $E(t)$ when 
compact flat FRW space-times are considered has been carried out, 
and we hope to publish our results shortly elsewhere. We anticipate, 
however, that the causes for such a behavior lie in the compactness 
of the manifold in at least one direction, on the one hand, 
and in the type of coupling between $Q$ and $\phi$, on the other 
hand.

A possible physical measure of the  degree of 
connectedness in these {\em compact\/} 3-manifolds can be made 
through the study of the number of emitted rays that return to 
the origin within a given lapse of time. 
According to this concept of degree of connectedness 
one learns from figures 2, 3 and 4 that the greater 
is the degree of connectedness the earlier is 
the occurrence of the first relative minimum in the energy 
function.
Incidentally, note that the extension of this concept 
of degree of connectedness to non-compact
3-manifolds implies that ${\cal R}^3$ has a null 
degree of connectedness. This, of course, is indicated
in figure 4, which shows a net exponential decay of the energy 
with the time; no relative minima and maxima come about, which 
means that no ray returns to the origin.

A simple inspection of the graphs for $E(t)$ clearly shows 
that the derivative 
$\dot{E} = \dot{Q}\,(\, \ddot{Q} + \Omega^2\, Q\,)$ 
of the energy function is discontinuous at a few points.  
Indeed, for the static case, for example, using (\ref{T1radeq}) 
one obtains
\begin{equation}
\dot{E}(t) = 2 \, \Gamma \, \dot{Q}(t) \left[ \, 
 \sum_{n_{1},n_{2},n_{3}} \frac{1}{a(\vec{n})}\,\,
 \Theta(t-a(\vec{n})) \,\, Q(t - a(\vec{n})\,) \,-\, 
		\dot{Q}(t) \right] \,.
\end{equation} 
From this equation one sees that the discontinuities occur
at $t=a(\vec{n})$, that is, they come about each time a new
term $ Q(t-a(\vec{n}))\,/\,a(\vec{n})$ is taken into account in
the right hand side of (\ref{T1radeq}).
A question which naturally arises here is whether the
inverse problem, i.e. that of determining the basic cell (topology)
corresponding to the spacelike $t=const$ sections from the graphs of
the energy $E(t)$, can be solved. Regarding this problem it is clear
that one can find the distances of the point-like source to its images  
by using where the discontinuities of $\dot{E}(t)$ take place, and 
the number of images at a given distance through the magnitude of the 
corresponding discontinuities. So, one can probe 
the topology of the 3-spaces at least in a few cases. 
It is not yet clear whether an algorithm for solving the inverse
problem for the most general (Euclidean) setting can be found,
though.

As far as we are aware~\cite{LachiezeReyLuminet1995}
this is the first work in which a physical consequence 
of adopting the flat hexagonal prisms  
${\cal H}_1$ and ${\cal H}_2$ has been studied.

It is of worth to emphasize that when the expansion of the 
universe is considered the degree of connectedness is less than 
the ones for the static cases. For the manifold ${\cal T}_{1}$ 
this can be seen by comparing figure 2 (static case 
$\dot{A}(t) = 0$) and figure 5 (monotone expansions 
$\dot{A}(t) > 0$). 

Before closing this article we should like to stress that
our study does not cover all possible spatially compact orientable 
flat FRW manifolds. Thus, for example, we have not 
considered that for the 3-manifolds ${\cal T}_1$ 
and ${\cal T}_3$ the basis of the basic cell need not
to be a square, it can be a parallelogram.
The restriction we have made, however, does not seem
to be decisive for the patterns of the behavior of 
the energy with the time we have found.              

To conclude we remark that the study of radiation damping in 
elliptic ($\kappa = 1$) FRW manifolds in which the $t=const$ 
sections are endowed with dif\/ferent (orientable compact) topologies 
is being carried out.

\vspace{5mm}

\section*{Acknowledgements}

We are grateful to an anonymous referee for valuable suggestions. 
We also thank the Brazilian scientific agencies CNPq and CAPES 
for financial support. 
A.B. is also grateful to CLAF.

\begin{figure} 
\caption{
A basic cell for the manifold ${\cal T}_4$ is a cube whose faces 
are pairwise identified as indicated in this figure. }
\label{fig1}
\end{figure}

\begin{figure}
\caption{
Behavior of the energy of the harmonic oscillator for $\Gamma=1$ and 
$\Omega^2 =30$ in static flat FRW space-times with ${\cal T}_1$ 
space slices. 
There are a few relative maxima followed by a growth of $E(t)$.
Two dif\/ferent ratios $h/a$ are considered ($h$ is the height 
and $a$ is the side of the square basis of the basic cell). } 
\label{fig2}
\end{figure}

\begin{figure}
\caption{
Behavior of the energy $E(t)$ of the harmonic oscillator
for $\Gamma=1$ and $\Omega^2 =30$
in static flat FRW space-times with 3-space ${\cal H}_1$.   
There are a few relative maxima followed by a growth of $E(t)$. 
It shows the energy {\it vs.} time curves for 
distinct ratios $h/a$ ($h$ is the height of the hexagonal prism 
and $a$ is shortest distance between two opposite sides of the 
regular hexagon). }
\label{fig3}
\end{figure}

\begin{figure}
\caption{
The time evolution of the energy of the harmonic oscillator
for flat, static FRW space-times with dif\/ferent topologies
for the spacelike $t=const\,$ sections: 
${\cal R}^3, \, {\cal T}_1$ and ${\cal H}_1$ 
(both with $a \, = \, h \, = \, 1$). 
Dif\/ferent degree of connectedness implies 
dif\/ferent patterns of $E(t)$. 
Here again $\Gamma=1$ and $\Omega^2 =30$. }
\label{fig4}
\end{figure}

\begin{figure}
\caption{
The time evolution of the energy of the harmonic oscillator
in FRW expanding space-times with 3-space
${\cal T}_1$ (for $a \, = \, h \, = \, 1$). Three types of 
dynamic expansion are shown:
linear [$\,\dot{A}(t) = 1\,$], square root [$\,A(t)= t^{1/2}\,$]
and inflationary 
[$\,\dot{A}(t) = \alpha \, e^{\alpha \,t} \,, \; \alpha^{-1} = 2^{1/2}\,$].
Here again $\Gamma=1$ and $\Omega^2 =30$. }
\label{fig5}
\end{figure}

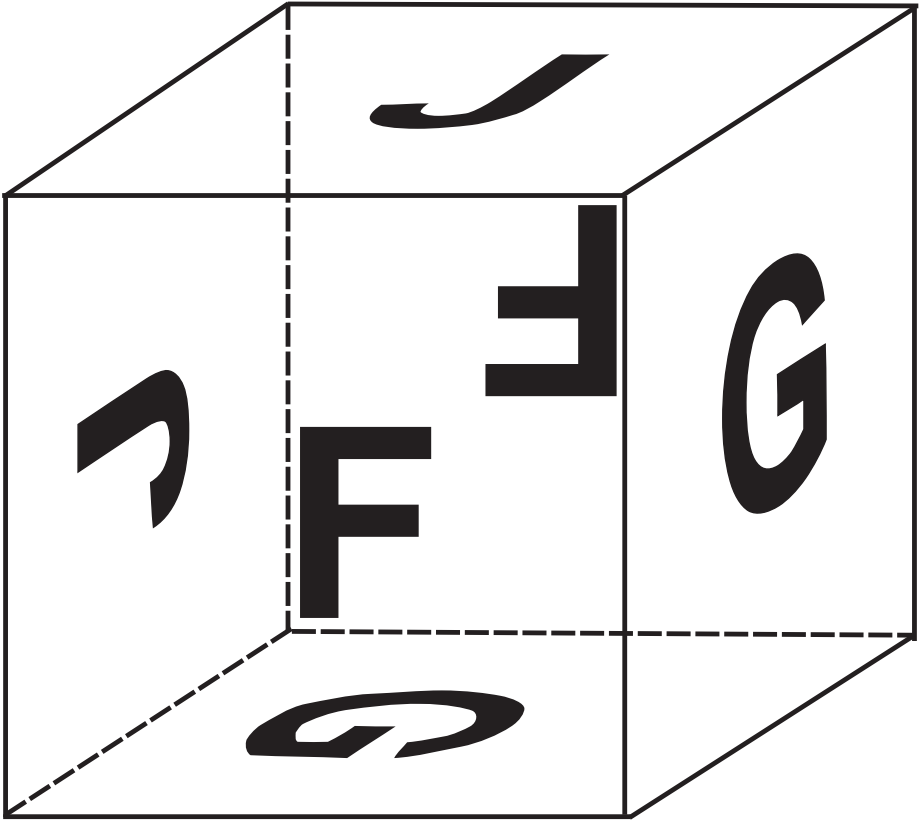
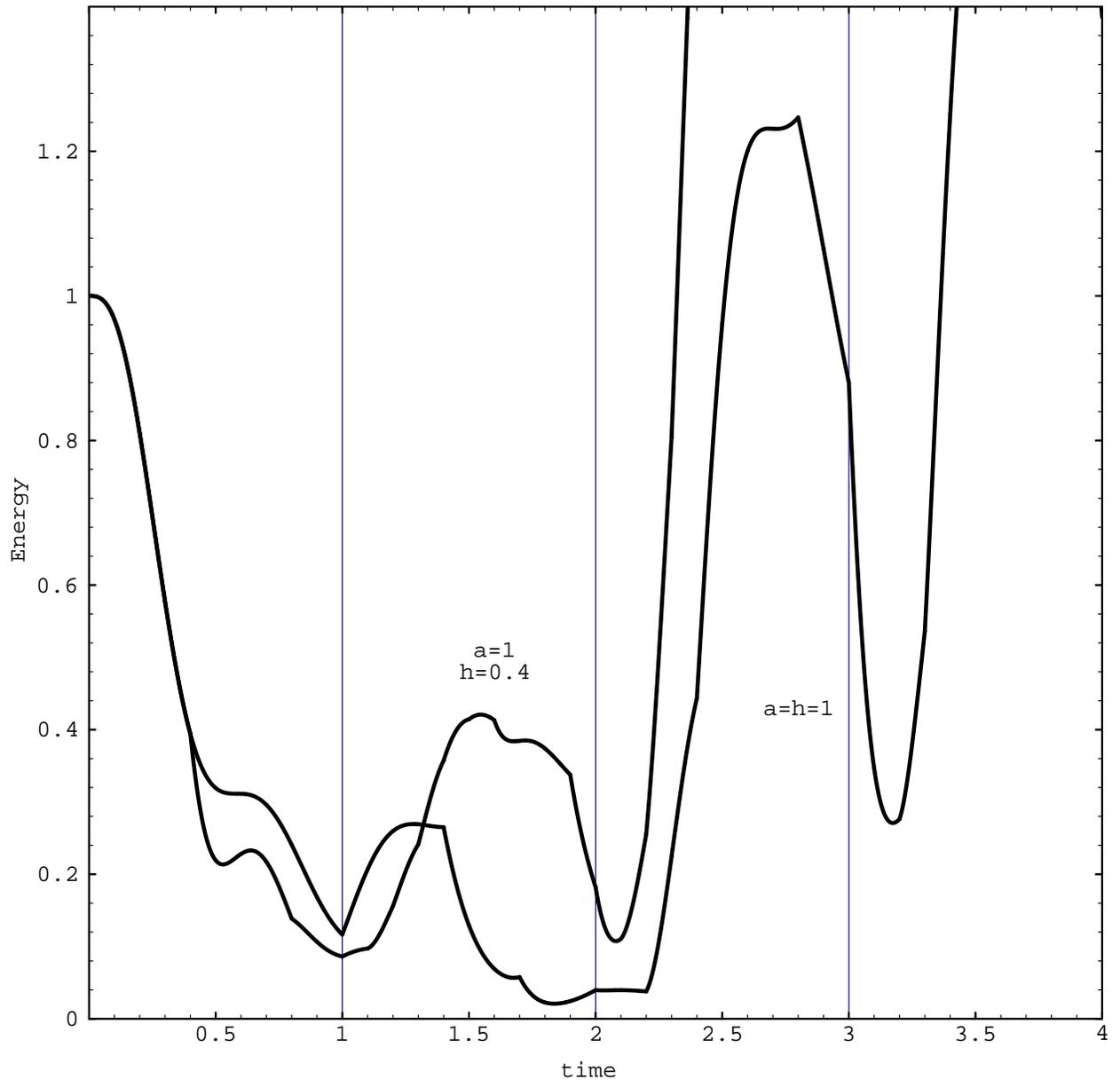
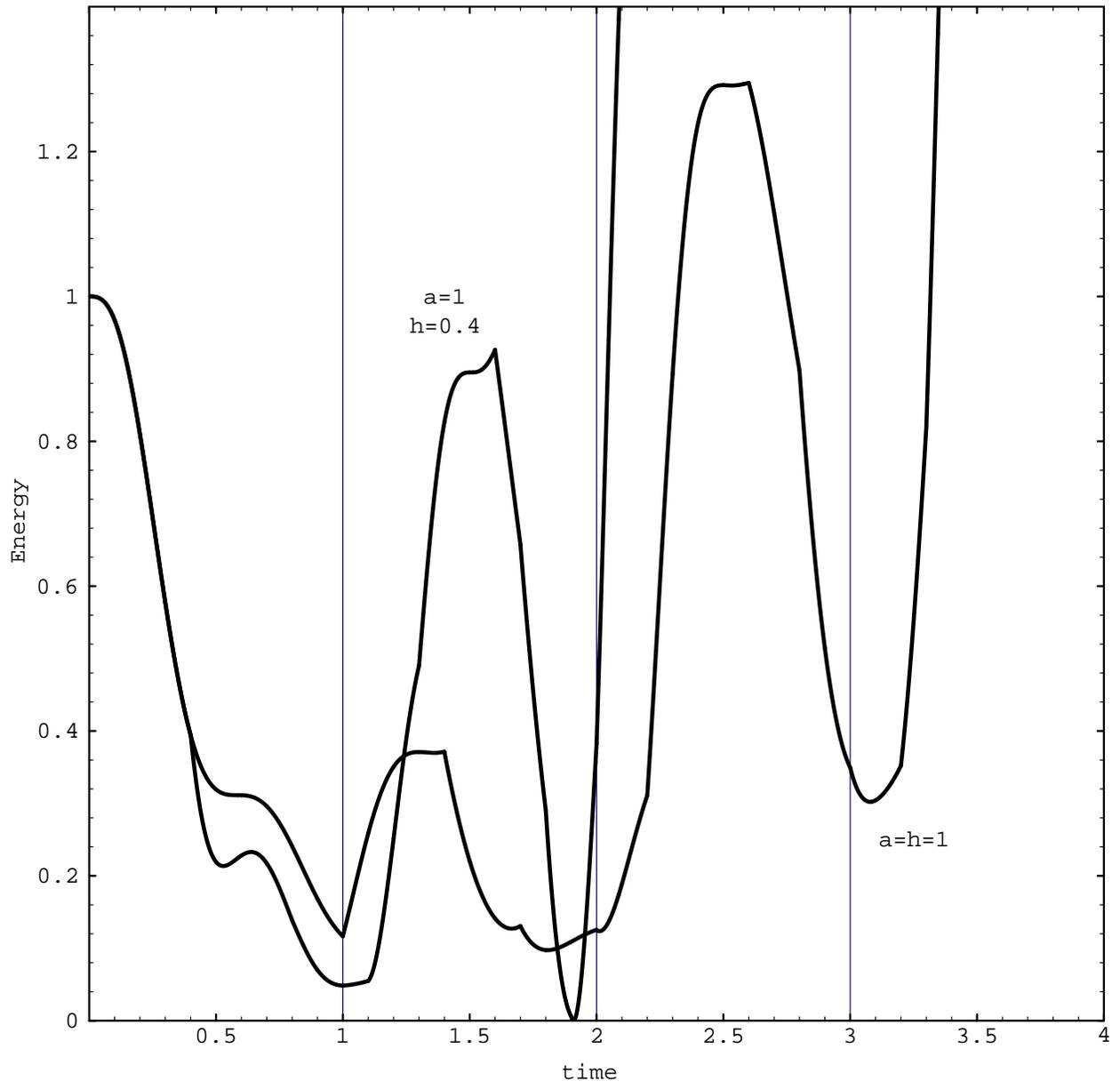
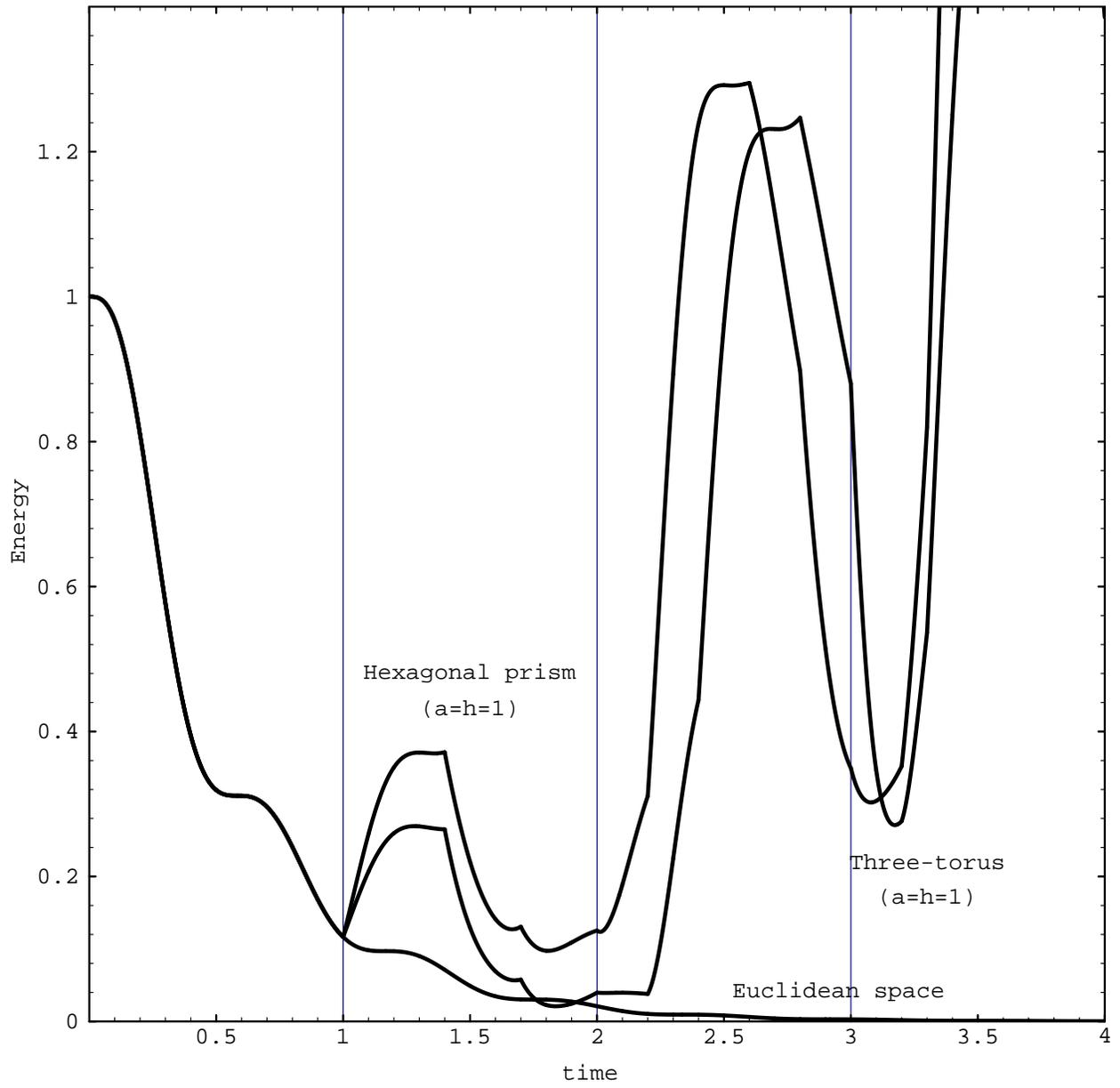
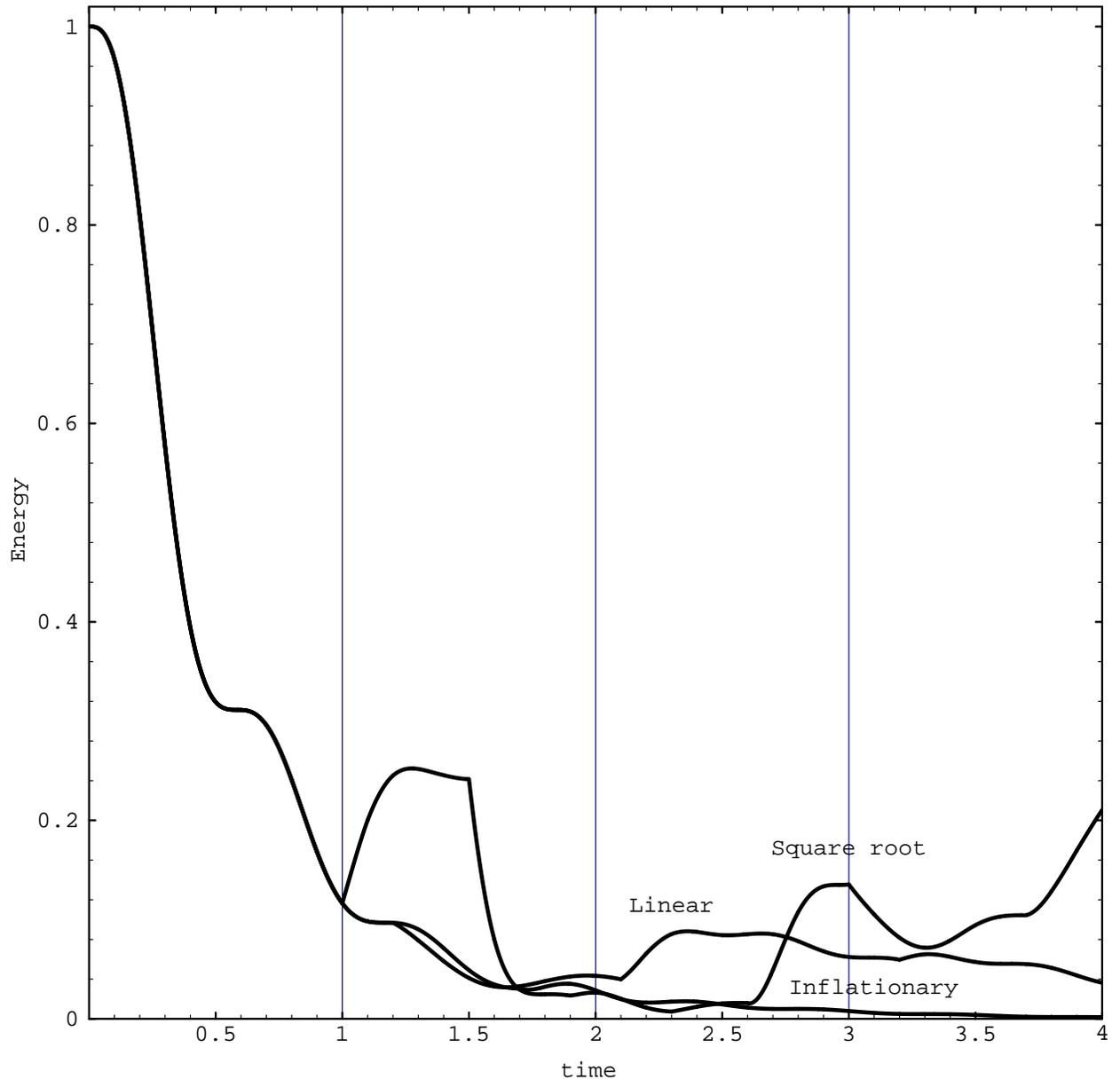
\end{document}